\documentclass
[
    a4paper,
    DIV=11,
    abstracton,
]
{scrartcl}

\usepackage[utf8]{inputenc}

\usepackage[inline]{enumitem}

\usepackage
{
    graphicx,
    amsmath,
    amssymb,
    indentfirst,
    xcolor,
    authblk,
    sectsty,
    array,
    xcolor,
    float,
    authblk,
    natbib
}

\usepackage[bf,normal]{caption}

\usepackage[pdffitwindow=false,
            plainpages=false,
            pdfpagelabels=true,
            pdfpagemode=UseOutlines,
            pdfpagelayout=SinglePage,
            bookmarks=false,
            colorlinks=true,
            hyperfootnotes=false,
            linkcolor=blue,
            urlcolor=blue!30!black,
            citecolor=green!50!black]{hyperref}

\graphicspath{{pics/}}

\allowdisplaybreaks

\DeclareMathOperator{\sorted}{sorted}

\title{\textbf{GraphKKE: Graph Kernel Koopman Embedding for Human Microbiome Analysis}}

\author[1]{Kateryna Melnyk\thanks{Correspondence: katerynam@zedat.fu-berlin.de}}
\author[1,2]{Stefan Klus}
\author[3]{\\Grégoire Montavon}
\author[1,4]{Tim O.F. Conrad}

\affil[1]{Department of Mathematics and Computer Science, Freie Universität Berlin}
\affil[2]{Department of Mathematics, University of Surrey}
\affil[3]{Electrical Engineering and Computer Science, Technische Universität Berlin}
\affil[4]{Zuse Institute Berlin}

\date{}

\begin{document}
\maketitle

\begin{abstract} 
More and more diseases have been found to be strongly correlated with disturbances in the microbiome constitution, e.g., obesity, diabetes, or some cancer types. Thanks to modern high-throughput omics technologies, it becomes possible to directly analyze human microbiome and its influence on the health status. Microbial communities are monitored over long periods of time and the associations between their members are explored. These relationships can be described by a time-evolving graph. In order to understand responses of the microbial community members to a distinct range of perturbations such as antibiotics exposure or diseases and general dynamical properties, the time-evolving graph of the human microbial communities has to be analyzed. This becomes especially challenging due to dozens of complex interactions among microbes and metastable dynamics. The key to solving this problem is the representation of the time-evolving graphs as fixed-length feature vectors preserving the original dynamics. We propose a method for learning the embedding of the time-evolving graph that is based on the spectral analysis of transfer operators and graph kernels. We demonstrate that our method can capture temporary changes in the time-evolving graph on both synthetic data and real-world data. Our experiments demonstrate the efficacy of the method. Furthermore, we show that our method can be applied to human microbiome data to study dynamic processes.
\end{abstract}

\section{Introduction}
\label{sec:intro}

Approximately every second cell in our body is a microbial cell. We are colonized by a diverse community of bacteria, archaea, and viruses, jointly referred to as the microbiome.  About 1.5 kg of microbes live almost everywhere on and in the human body as symbionts, e.g., on the skin, in the mouth, or in the gut. They have a strong influence on both their hosts and environments. For example, more and more diseases have been found to be strongly correlated with the disturbances in the microbiome constitution, e.g., obesity \citep{ObesityMicrobiome1, ObesityMicrobiome2, ObesityMicrobiome3}, diabetes \citep{DiabetesMicrobiome1}, or some cancer types \citep{CancerMicrobiome1, CancerMicrobiome2}. Furthermore, recent studies have revealed that gut microbiome also has a huge impact on brain functions and is related to disorders such as Alzheimer's disease \citep{AlzheimerMicrob}. Most studies aiming at understanding the differences in the microbiome profiles of healthy and ill individuals, however, are focused on statistical constitution analysis, omitting the large variety of complex microbe--microbe and host--microbe interactions, which can be modeled as time-evolving graphs. 

\begin{figure}
    \centering
    \includegraphics[width=0.9\textwidth]{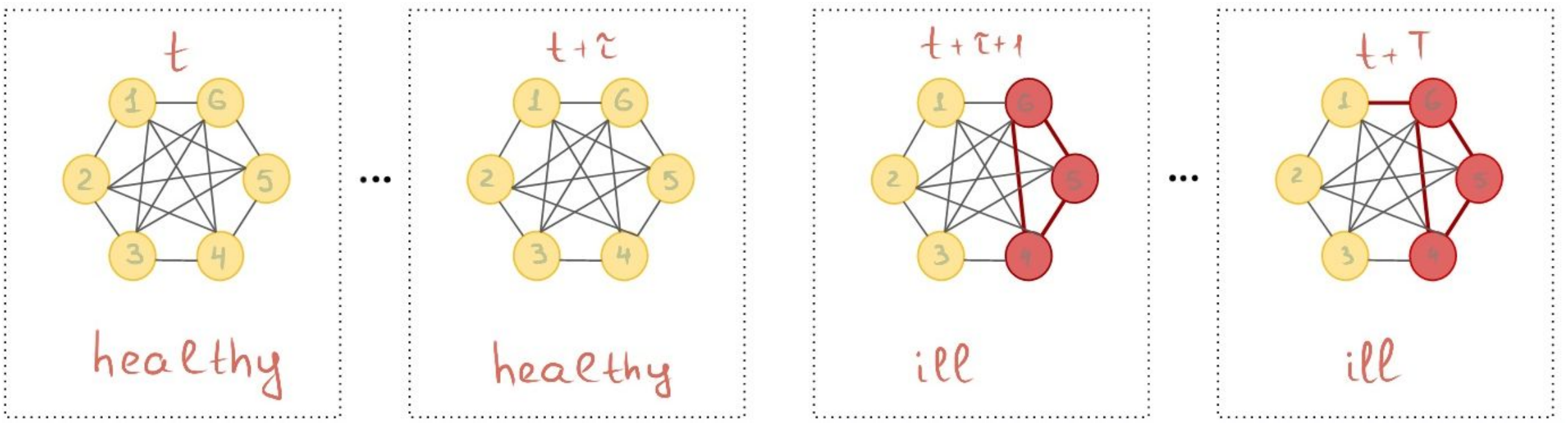}
    \caption{An example of a time-evolving graph of microbe interactions with two metastable states: \emph{healthy} and \emph{ill}. Red color of vertices means that the concentrations of microbes decreased after a person became ill at time $t + \tau + 1$. This reduction results in the change of the topology of the time-evolving graph characterized by removing the edges between vertices in red and some vertices in yellow.}
    \label{fig:figure1}
\end{figure}

It has also been found that although the constitution of the microbiome is constantly changing throughout our lives (in response to environmental factors), a healthy human microbiome can be considered as a metastable state lying in a minimum of some ecological stability landscape \citep{microbiomeState}.  Broadly speaking, metastability can be observed when for short timescales, the system appears to be equilibrated, but at larger time scales, undergoes some transitions from one metastable state to other metastable states \citep{Bovier06:metastability}. This phenomenon occurs in dynamical systems of various structures, including systems with vector-valued states, but also systems represented as time-evolving graphs. In this context, metastability means that the graph structure is stable for a relatively long time (up to small perturbations) before the system undergoes a critical transition --- e.g., when it reaches a tipping point~--- and shifts to a different metastable state.

As an illustration of a time-evolving graph that lies in an energy landscape with two metastable states, consider the time-evolving microbiome interaction graph shown in Figure \ref{fig:figure1}, where vertices represent the concentrations of bacteria species and edges pairwise associations between them. In this example, a disease can be thought of as a perturbation that displaces the microbiome composition from its equilibrium (\emph{healthy}) state. The consequence of this displacement is the reduction of the concentration in the red vertices and the removal of edges that connect red vertices. Given an evolution of the graphs (in this example, the evolution of the microbe interactions), we aim at analyzing dynamics occurring in the graph over time, namely, extracting the number of metastable states and their locations, substructures of a graph, which characterize the state space (e.g., the difference in the microbe interactions between the states \emph{healthy} and \emph{ill}). Moreover, the detection of the metastable states in the time-evolving graph can serve additional purposes such as graph clustering. 

 \begin{figure}
    \centering
    \includegraphics[width=0.8\textwidth]{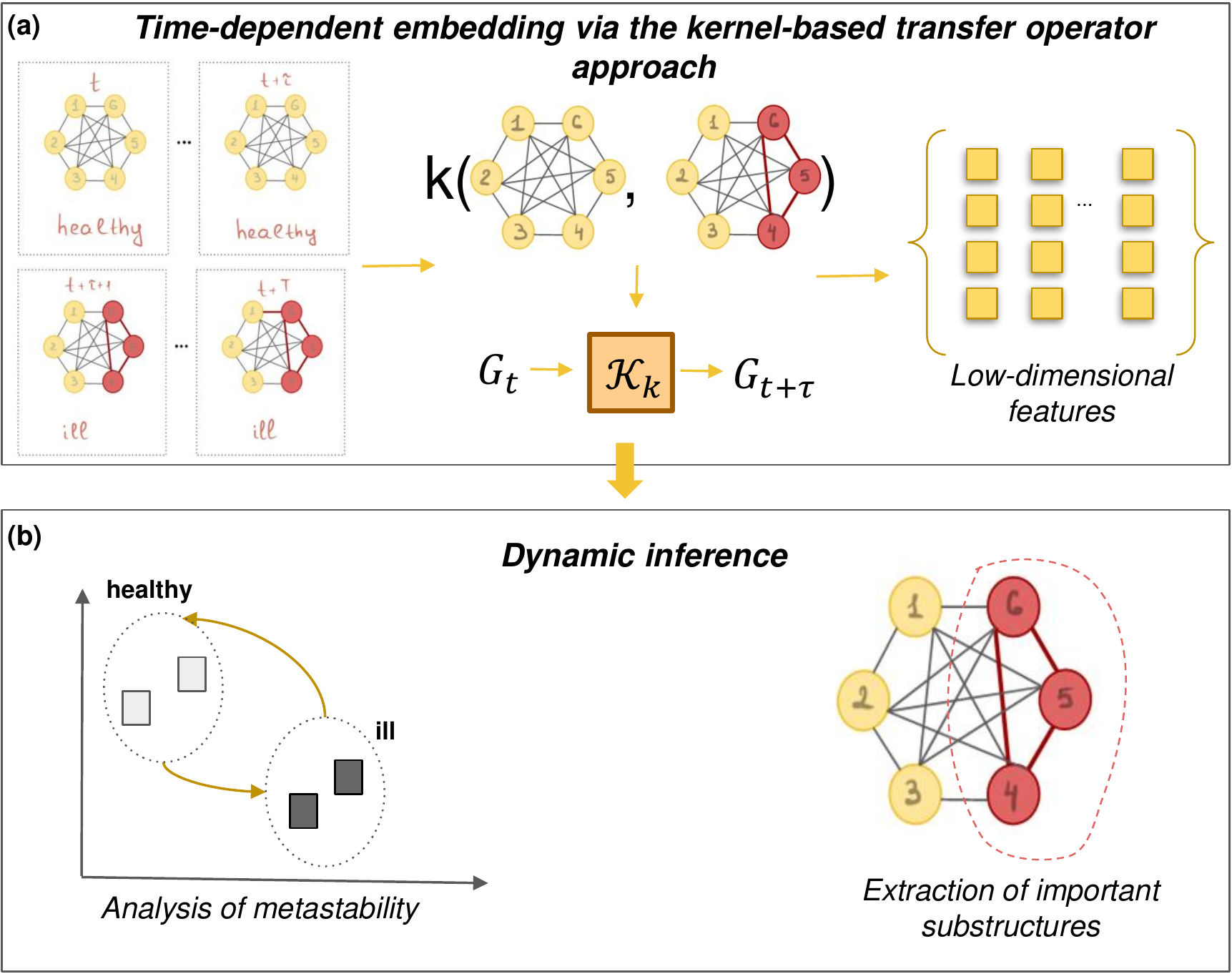}
    \caption{An illustration of the proposed method and challenges which we aim to overcome. \textbf{(a)} Learning transfer operators using graph kernels, where $k(\cdot, \cdot)$ is a graph kernel and $\mathcal{K}_k$ is the Koopman operator. \textbf{(b)} In the learned embedding space it is possible to detect metastable states and to determine distinct substructures.}
    \label{fig:graphical_abst}
\end{figure}

\paragraph*{\textbf{Related work.}} Two potential ways to detect metastable states in a time-evolving graph (e.g., the states \emph{healthy} and \emph{ill} in our example) are the following: 
\begin{enumerate}[topsep=-\parskip,itemsep=-\parsep,leftmargin=2.5em,label=\arabic*.]
    \item A typical solution would be to analyze the time-evolving graph directly in the space of graphs without taking into account potential temporal correlations. Practically, this can take the form of a simple kernel-based graph clustering algorithm. Classic graph kernels decompose graphs into substructures (e.g., walks \citep{RandomWalk}, subgraphs \citep{Graphlet}, paths \citep{Shortest_path}, and subtrees \citep{WL}) and count the number of common substructures between graphs in order to obtain the feature vectors. Afterwards, these feature vectors can be used by various machine learning approaches to cluster snapshots of the time-evolving graphs. The problem with such methods is that they are incapable of capturing the time-information, which is crucial for time-evolving graphs with metastability.
    \item Another possible way is graph representation learning, which aims at finding a mapping that embeds the system into some low-dimensional space. That is, we represent a single snapshot of the time-evolving graph at each time point by a single vector retaining the original properties of the dynamics. After finding the optimal embedding space, the low-dimensional representation can be used as a feature input for diverse machine learning approaches for analyzing time-series data.
\end{enumerate}

The recently proposed methods for graph representation learning focus mostly on static graphs. These methods can be broadly divided into two categories. The first category comprises methods for embedding graph substructures (e.g., vertices or subgraphs), see \citet{Deepwalk, node2vec, SDNE, HOPE}. For instance, DeepWalk \citep{Deepwalk} and node2vec \citep{node2vec} are approaches that use random walks to produce embeddings. The only difference between them is that node2vec utilizes two hyperparameters, where one of them controls the likelihood of a random walk to return to the previously visited vertex and another parameter controls the likelihood to explore undiscovered parts of a graph. DeepWalk first traverses the graph with random walks in order to extract local structures and then it uses the Skip-Gram algorithm to learn embeddings. The second category pertains to representation learning of the entire graph, which is used for the classification/clustering of the set of graphs. The graph2vec approach \citep{graph2vec} learns the embedding of the set of graphs using the idea of the Skip-Gram from doc2vec \citep{doc2vec}. It comprises two main components: (1) The generation of rooted subgraphs around every vertex using the Weisfeiler--Lehman relabeling process from \citet{WL}; (2) Learning the embedding of the given graphs following the Skip-Gram with negative sampling procedure. Although this approach is capable of projecting the entire set of graphs into low-dimensional space, it does not capture the time-evolution of the graph.

Recently, some work has also been done on learning the embedding vectors of vertices in the time-evolving graph. Dyngraph2vec \citep{dyngraph2vec} is a deep-learning based approach which learns both the topological patterns in a graph and the temporal transitions using multiple nonlinear layers and recurrent layers. Moreover, it uses the lookback hyperparameter in the recurrent layers to control the length of temporal patterns. The idea of DynamicTriad \citep{dynTriad} is to use a group of three vertices, a so-called \emph{triad}, to model the dynamic changes of graph structures. This approach only considers patterns within two time steps, which means that it cannot capture patterns that exist for a longer period of time. The main disadvantage of the substructure representation learning approaches, both for static and for time-evolving graphs, is that they are not able to project the entire set of snapshots of the time-evolving graph into low-dimensional space.

\paragraph*{\textbf{Contribution.}} To this end, we present an approach named \textbf{graphKKE} (the overall structure is shown in Figure \ref{fig:graphical_abst}), which is, to our knowledge, the first approach for representation learning of an entire time-evolving graph. Inspired by the proposed kernel transfer operator approach for molecular conformation analysis \citep{Klus2017EigendecompositionsOT, KlusMolecularAnalysis}, we use the same approach for learning the embeddings of time-evolving graphs. The method is based on the spectral analysis of transfer operators, such as the Perron--Frobenius or Koopman operator in a reproducing kernel Hilbert space.

Overall, we highlight the following contributions: 
\renewcommand\labelitemii{$\square$}
\begin{itemize}
    \item[$\square$] We propose \textbf{graphKKE}, a novel unsupervised representation learning technique to analyze a time-evolving graph, i.e., class labels of the graphs are not required for learning their embedding. Moreover, we demonstrate the applicability of the graph kernels to time-evolving graphs. Our method is not only capable of preserving the information about the underlying dynamical graph patterns but also of taking into account the topological structure of the graph.
    \item[$\square$] We present a new simulation method for constructing artificial benchmark datasets of time-evolving graphs with metastability and with graph structures of different complexity. We demonstrate that \textbf{graphKKE} significantly outperforms other methods for graph representation learning on several benchmark problems.
    \item[$\square$] We illustrate that \textbf{graphKKE} can extract the important associations among microbes and capture the temporal changes occurring in the time-evolving microbiome interaction graph.
\end{itemize}

The remainder of this paper is organized as follows: In Section \ref{sec: problem_state}, the problem of learning the embeddings of time-evolving graphs with metastable behavior is defined. In Section \ref{sec: method}, we introduce transfer operators, graph kernels, and the method for the approximation of transfer operators using graph kernels. A model for the simulation of time-evolving benchmark graphs with metastability and the experiments with these benchmark datasets are presented in Sections \ref{sec:data} and \ref{sec:experiments}. Eventually, Section \ref{sec:microbiome_d} illustrates that it is possible to obtain a meaningful low-dimensional representation for microbiome data.

\section{Problem statement}
\label{sec: problem_state}

In order to state the problem formally, let us first  introduce the necessary notations and definitions.

A graph $G$ is a pair $(V, E)$ with a non-empty set of vertices $V(G)$ and a set of edges $E(G) = \{(v_i, v_j) \mid v_i, v_j \in V\}$. The set $V(G)$ often represents the objects in the data and $E(G)$ relations between objects. We define \textit{the adjacency matrix} of the graph $G$ as the $n \times n$ matrix $A$ with $A_{ij} = 1$ if the edge $(v_i, v_j) \in E(G)$, and 0 otherwise. Furthermore, we say that $\bar{G} = (\bar{V}, \bar{E})$ is \textit{a subgraph} of a graph $G = (V, E)$ if and only if $\bar{V} \subseteq V$ and $\bar{E} \subseteq E \wedge ((v_i, v_j) \in \bar{E} \Rightarrow v_i, v_j \in \bar{V})$.

Given a time-evolving graph $\mathbb{G}$ as a sequence of $T$ graphs $\mathbb{G} = (G_0,\dots, G_{T-1})$ at the consecutive time points $\{0,\dots, T-1\}$ for some $T \in \mathbb{N}$. We call $G_t$ a time-snapshot of $\mathbb{G}$ at time $t$. We focus in particular on metastability properties of the time-evolving graph, that is, the property of being stable for a long time, and occasionally undergoing critical transitions from one state to another state, with a significant change in the edges and/or nodes. More formally, we say that the time-evolving graph $\mathbb{G}$ exhibits {\em metastable} behavior if $\mathbb{G}$ can be partitioned into $s$ subsets $\mathbb{G} = \mathbb{G}_0 \cup \dots \cup \mathbb{G}_{s-1}$ for some $s \ll T$ such that for each time point $t \in \{0,\dots, T-1\}$
\begin{equation*}
    P(G_{t + 1} \in \mathbb{G}_i \mid G_t \in \mathbb{G}_j) \ll 1, \text{ if } i \neq j
\end{equation*}
and 
\begin{equation*}
    P(G_{t + 1} \in \mathbb{G}_i \mid G_t \in \mathbb{G}_j) \approx 1, \text{ if } i = j.
\end{equation*}
We call $\mathbb{G}_0, \dots, \mathbb{G}_{s-1}$ metastable states of the time-evolving graph $\mathbb{G}$ and each $G_t, t = 0, \dots, T-1$, belongs to exactly one of the states $\mathbb{G}_i$. In most cases, each state $\mathbb{G}_i$ is characterized by a certain pattern of graph attributes (i.e., edges, vertex labels).

We define our problem as follows: \textit{Given a time-evolving graph $\mathbb{G} = (G_0, \dots, G_{T - 1})$ with assumed metastable behavior, we aim to represent each time-snapshot $G_t$ as a vector in a low-dimensional space $\mathbb{R}^m$, where  $m$ is a number of embedding dimensions, retaining the metastable behavior of $\mathbb{G}$.} 

Commonly, the number of embedding dimensions $m$ is a hyperparameter that has to be tuned in order to obtain a good performance, in our approach we will show that the number of embedding dimensions $m$ can be chosen to be the number of states $s$, which eliminates the need to optimize this hyperparameter. 

\section{GraphKKE: Graph Kernel Koopman Embedding}
\label{sec: method}
In what follows, we first introduce transfer operators, kernel functions, and graph kernels. Afterwards, we present our approach --- \textbf{graphKKE} --- that is capable of learning embeddings of time-evolving graphs preserving temporal changes in a low-dimensional space. 

\subsection{\large{Transfer operators}}
In order to capture the temporal changes in the time-evolving graph, transfer operator theory will be used in our method. Therefore, we will briefly discuss transfer operators and their applicability in the analysis of dynamical systems (for details, see \citet{NumApprTranferOperators}). Information about the evolution of the system is contained in the spectral properties (such as eigenvalues and eigenfunctions) of linear operators. The most commonly used examples of such operators are the Koopman operator and the Perron--Frobenius operator. 

Let $\{X_t\}_{t\geq 0}$ be a stochastic process defined on a high-dimensional state space $\mathbb{X} \subset \mathbb{R}^d$. The pointwise evolution of $X_t$ can be formally described by the transition density function $p_\tau(y \mid x)$, which gives the probability to find the process at a point $y$ after some lag time $\tau$, given that it started in $x$ at time 0. More formally, the transition density function is
\begin{equation*}
    p_\tau(y \mid x) = P(X_{t + \tau} = y \mid X_{t} = x).
\end{equation*}

With the aid of the transition density function, the Koopman operator expresses the evolution of a function of the state, also called observable, whereas the Perron--Frobenius operator evolves probability densities. Let $f_t \in L^\infty(\mathbb{X})$ be an observable of the system. Then the Koopman operator $\mathcal{K}_{\tau} \colon L^\infty(\mathbb{X}) \rightarrow L^\infty(\mathbb{X})$ is defined by
\begin{equation}
    \mathcal{K}_{\tau}f_t(x) = \int p_{\tau}(y \mid x)f_t(y)dy.
    \label{eqn:koopm_op}
\end{equation}

The evolution of probability densities can be described in a similar way. Assume the initial density of the system is given by $g_t \in L^1(\mathbb{X})$. Then the Perron--Frobenius operator $\mathcal{P}_{\tau} \colon L^1(\mathbb{X}) \rightarrow L^1(\mathbb{X})$ is defined by
\begin{equation*}
    \mathcal{P}_{\tau}g_t(x) = \int p_{\tau}(x \mid y)g_t(y)dy.
\end{equation*}

A density $\pi$ is called \emph{invariant density} or \emph{equilibrium density} if it is invariant under the action of $\mathcal{P}_\tau$, that is, $\mathcal{P}_\tau\pi = \pi$. Let $u_t(x) = \pi(x)^{-1}g_t(x)$ be a probability density with respect to the equilibrium density $\pi$. Then, the Perron--Frobenius operator with respect to the equilibrium density is defined as

\begin{equation*}
    \mathcal{T}_\tau u_t(x) =\frac{1}{\pi(x)} \int p_{\tau}(x \mid y)\pi(y)u_t(y)dy.
\end{equation*}

Both the Koopman operator $\mathcal{K}_{\tau}$ and the Perron--Frobenius operator $\mathcal{P}_{\tau}$ are linear, infinite-dimensional operators, which are adjoint to each other and, therefore, it should not matter which one we choose to study the behavior of the system. Moreover, although they are typically defined on the function spaces $L^1$ and $L^{\infty}$, we assume that the operators are well-defined on $L^2$ (for details, see \citet{NumApprTranferOperators}).
    
The information about the long-term behavior of the dynamical system is encoded in the spectral properties of these operators such as eigenvalues and eigenfunctions \citep{Klus2017EigendecompositionsOT}. More precisely, eigenfunctions with eigenvalues close to 1 of both Koopman and Perron--Frobenius operators contain information about the locations of metastable states in the state space $\mathbb{X}$.

Since transfer operators are infinite-dimensional, the goal is to obtain a finite-dimensional approximation of these operators. Below, we will show how to obtain a finite-dimensional approximation of transfer operators utilizing the evaluation of graph kernels on training data.


\subsection{\large{Graph kernels}}
\label{sec: graph_kernel}
In this section, we describe kernel functions and a neighborhood aggregation graph kernel, the $1$-dimensional Weisfeiler--Lehman kernel, since all our experiments make use of this graph kernel. However, one can potentially use other graph kernels, which can be tailored to specific applications. 

\paragraph*{\textbf{Kernel function.}} 
Kernel-based methods are machine learning algorithms that learn by comparing any pair of data points using similarity measures called kernel functions. We will say that $k \colon \mathbb{X} \times \mathbb{X} \rightarrow \mathbb{R}$ is a kernel on $\mathbb{X}$ if there is a Hilbert space $\mathbb{H}$ and a feature map $\varphi \colon \mathbb{X} \rightarrow \mathbb{H}$ such that
\begin{equation}
    k(x, x') = \langle \varphi(x), \varphi(x') \rangle
\end{equation}
for $x, x' \in \mathbb{X}$ and where $\langle \cdot , \cdot \rangle$ is the inner product on $\mathbb{H}$.
A feature map $\varphi$ exists if and only if $k$ is a positive-semidefinite function. However, the kernel is normally not defined by an explicit representation of $\varphi$, but instead, each kernel implicitly defines a potentially infinite-dimensional mapping $\varphi$.

For a given set of data points $x_0, \dots, x_m \in \mathbb{X}$, the matrix $K$ with $K_{ij} = k(x_i, x_j)$ for $i, j = 0, \dots, m$, is called \emph{Gram matrix}. The Gram matrix is positive semidefinite for all possible $\{x_0, \dots, x_m\}$.

Now let $\mathbb{G}$ be a sequence of graphs, then a kernel $ k \colon \mathbb{G} \times \mathbb{G} \rightarrow \mathbb{H}$ is called a graph kernel.

\paragraph*{\textbf{Gaussian kernel.}}
The most popular kernel function used in numerous kernel-based methods is the Gaussian kernel, which for two graphs $G$ and $\widehat{G}$ can be defined as
\begin{equation*}
    k(G, \widehat{G}) = \exp \Big(-\frac{\|A - \widehat{A}\|^2}{2 \sigma^2} \Big),
\end{equation*}
where $A$ and $\widehat{A}$ are the respective adjacency matrices, $\sigma > 0$ is the bandwidth parameter, and $\|\cdot\|$ the Frobenius norm. The Hilbert space $\mathbb{H}$ spanned by the Gaussian kernel is an infinite-dimensional space. Furthermore, it can be shown that continuous functions on a bounded domain can be approximated arbitrarily well by (weighted sums of) Gaussian kernels. For polynomial kernels, for instance, this is not the case.

\paragraph*{\textbf{Weisfeiler--Lehman kernel.}} In this work, we will use a neighborhood aggregation kernel --- the Weisfeiler--Lehman (WL) kernel \citep{WL} --- for graphs with discrete vertex labels. However, one could choose any other class of graph kernels such as \emph{graphlet kernels} from \citet{Graphlet} or \emph{random walk kernels} from \citet{RandomWalk}.

We will briefly give an overview of the Weisfeiler--Lehman kernel. Let $G$ and $\widehat{G}$ be graphs and $l^{(0)}$ be a set of unique original vertex labels of $G$ and $\widehat{G}$. The key idea of this kernel is to augment each vertex label by the sorted set of neighboring vertex labels, and then to compress the augmented label into some new label using a hash function $f$. That is, at each iteration $h = 1, \dots$, the 1-dimensional Weisfeiler--Lehman kernel computes a new set of vertex labels $l^{(h)}$ such that
\begin{equation*}
    l^{(h)}_v = f\Big(l^{(h-1)}_v +(l^{(h-1)}_{u_0} + ... + l^{(h-1)}_{u_k})\Big),\\
    \{u_0, ..., u_k\} \in \sorted(\mathcal{N}(v)),
\end{equation*}
$\forall v \in V(G) \cup V(\widehat{G})$ and where the symbol ``+'' denotes the concatenation of strings, $\mathcal{N}(v)$ the set of neighbors of a vertex $v$, and $\sorted(\mathcal{N}(v))$ means that vertex labels need to be sorted before concatenation. The hash function $f$ is chosen in such a way that  $f(l^{(h)}(v)) = f(l^{(h)}(v^\prime))$ if and only if $l^{(h)}(v) = l^{(h)}(v^\prime)$, $v, v^\prime \in V(G) \cup V(\widehat{G})$. The next step is to compute a feature vector for each graph $G$ and $\widehat{G}$ at each iteration $h$:
\begin{equation*}
    \varphi^{(h)}(G) = (C^{(h)}(G, l^{(h)}_0), ...,C^{(h)}(G, l^{(h)}_{|l^{(h)}|})), 
\end{equation*}
where $l^{(h)} = \{l^{(h)}_0, l^{(h)}_1, \dots, l^{(h)}_{|l^{(h)}|}\}$ denotes the set of compressed vertex labels at iteration $h$ and $C^{(h)}(G, l^{(h)}_i)$ is the number of occurrences of a label $l^{(h)}_i$ in the graph $G$ at iteration $h$.

Finally, the Weisfeiler--Lehman kernel for two graphs $G$ and $\widehat{G}$ is defined as:
\begin{equation*}
    k(G, \widehat{G}) = \langle \varphi^{(0)}(G), \varphi^{(0)}(\widehat{G}) \rangle + ... + \langle \varphi^{(h)}(G), \varphi^{(h)}(\widehat{G}) \rangle.
\end{equation*}
We chose the WL kernel because it outperformed other kernels in terms of runtime in our experiments. According to \cite{WL}, the WL subtree kernel on a pair of graphs can be computed in time $O(hm)$, where $h$ is the number of iterations and $m$ the number of edges, whereas the random walk kernel \citep{RandomWGartner} on a pair of graphs has the runtime complexity $O(n^6)$, where $n$ is the number of nodes. Moreover, it is also competitive in terms of accuracy with state-of-the-art kernels. But, as mentioned above, one could use other graph kernels as well. The optimal choice depends strongly on the dataset.

In the next subsection, we will introduce an approach for learning the embedding of a time-evolving graph using transfer operators and graph kernels. 

\subsection{\large{Method overview --- graphKKE}}
\label{sec: graphKKE}

Now, we introduce a graph kernel-based approximation method for time-evolving graphs inspired by the method proposed in \citet{KlusMolecularAnalysis}.

Since we cannot compute eigendecompositions of infinite-dimensional operators numerically, typically suitable finite-dimensional subspaces are considered. It was shown that the initial eigenvalue problem on $L^2$ can be approximated by an eigenvalue problem defined on the reproducing kernel Hilbert space $\mathbb{H}$ utilizing only kernel evaluations.

Assume we have measurement data, given by a time-evolving graph $\mathbb{G} = (G_0, ...,G_{T-1})$, where each $G_t$ is a single snapshot of $\mathbb{G}$ at time point $t$ and $\widehat{\mathbb{G}}$ is a set of graphs mapped forward for a time lag $\tau$, that is, $\widehat{G}_t = G_{t+\tau}$.

It was shown in \citet{Klus2017EigendecompositionsOT} that in order to find eigenfunctions of transfer operators, we need to solve auxiliary matrix eigenvalue problems, given by
\begin{equation}
\label{eqn: koopman_appr}
    K_{\mathbb{G}\mathbb{G}}^{-1} K_{\widehat{\mathbb{G}} \mathbb{G}} \tilde{\phi} = \lambda\tilde{\phi}
\end{equation}
and 
\begin{equation}
\label{eqn: perron_appr}
    K_{\mathbb{G}\mathbb{G}}^{-1} K_{\mathbb{G}\widehat{\mathbb{G}}} \tilde{\phi} = \lambda \tilde{\phi},
\end{equation}
where \([K_{\mathbb{G}\mathbb{G}}]_{ij} = k(G_{i}, G_{j})\), \([K_{\widehat{\mathbb{G}}\mathbb{G}}]_{ij} = k(\widehat{G}_{i}, G_{j})\) denote Gram matrices, $k(\cdot, \cdot)$ is a graph kernel, and $K_{\mathbb{G}\widehat{\mathbb{G}}} = K_{\widehat{\mathbb{G}}\mathbb{G}}^\top$. The equations \eqref{eqn: koopman_appr} and \eqref{eqn: perron_appr} approximate the Koopman operator and Perron--Frobenius operator, respectively. 

This eigenvalue problem is closely related to kernel canonical correlation analysis (kernel CCA), see \citet{Klus2019KernelCC}. Kernel CCA computes eigenfunctions of the forward-backward dynamics to identify so-called coherent sets. Coherent sets are a generalization of metastable sets and are regions of the state space that are not distorted over a certain time interval.

Additionally, in order to evaluate the eigenfunctions of these operators at a given graph, we set
\begin{equation*}
    \phi = \Psi \tilde{\phi},
\end{equation*}
if $\tilde{\phi}$ is the solution of the eigenvalue problem \eqref{eqn: koopman_appr}.

Otherwise, if $\tilde{\phi}$ is the solution of the eigenvalue problem \eqref{eqn: perron_appr}, we set
\begin{equation*}
    \phi = \Psi K^{-1}_{\mathbb{G}\mathbb{G}} \tilde{\phi}, 
\end{equation*} 
where $\Psi = [k(\cdot, G_0), \dots, k(\cdot, G_{T - 1})]$ is called a feature matrix.

We assume that $K_{\mathbb{G}\mathbb{G}}$ is non-singular or otherwise we replace the inverse by its regularized version $(K_{\mathbb{G}\mathbb{G}} + \eta I)^{-1}$, where $\eta\ge0$ is a ridge parameter. This regularization is known as Tikhonov regularization.

Furthermore, if $k(\cdot, \cdot)$ is a graph kernel, then we apply the following normalization:
\begin{equation*}
    k_{norm}(G_i, G_j) = \frac{k(G_i, G_j)}{\sqrt{k(G_i, G_i) \, k(G_j, G_j)}},
\end{equation*}
for all $i, j = 0, \dots, T-1$. The same normalization is applied to graphs in both $\mathbb{G}$ and $\widehat{\mathbb{G}}$.

The number of states $s$ in the time-evolving graph $\mathbb{G}$ is determined by the number of dominant eigenvalues close to 1. That is, if we have $s$ dominant eigenvalues close to 1, then the time-evolving graph can be divided into $s$ subsets $\mathbb{G} = \mathbb{G}_0 \cup \dots \cup \mathbb{G}_{s-1}$. Moreover, all information about long-term behavior of the time-evolving graph $\mathbb{G}$ is contained within the eigenfunctions associated with $s$ dominant eigenvalues close to 1. 
All things considered, the dominant eigenvalues can be used to determine the number of states $s$ in the data and the dimension of a new low-dimensional space. The eigenfunctions associated with the dominant eigenvalues close to 1 are considered as a low-dimensional representation of the time-evolving graph $\mathbb{G}$.

\section{Generating benchmark data with metastability} 
\label{sec:data}

Most of the benchmark data sets such as those from chemo- and bio-informatics domains, see \citet{KKMMN2016}, can be represented by static graphs. Thus, these datasets are not appropriate for our purposes, since they do not have time information and metastable behavior. Hence, in this section we present a model for generating time-evolving graphs with a comprehensible structure to estimate the performance of the proposed method.

\begin{figure}
    \centering
    \includegraphics[width=0.7\textwidth]{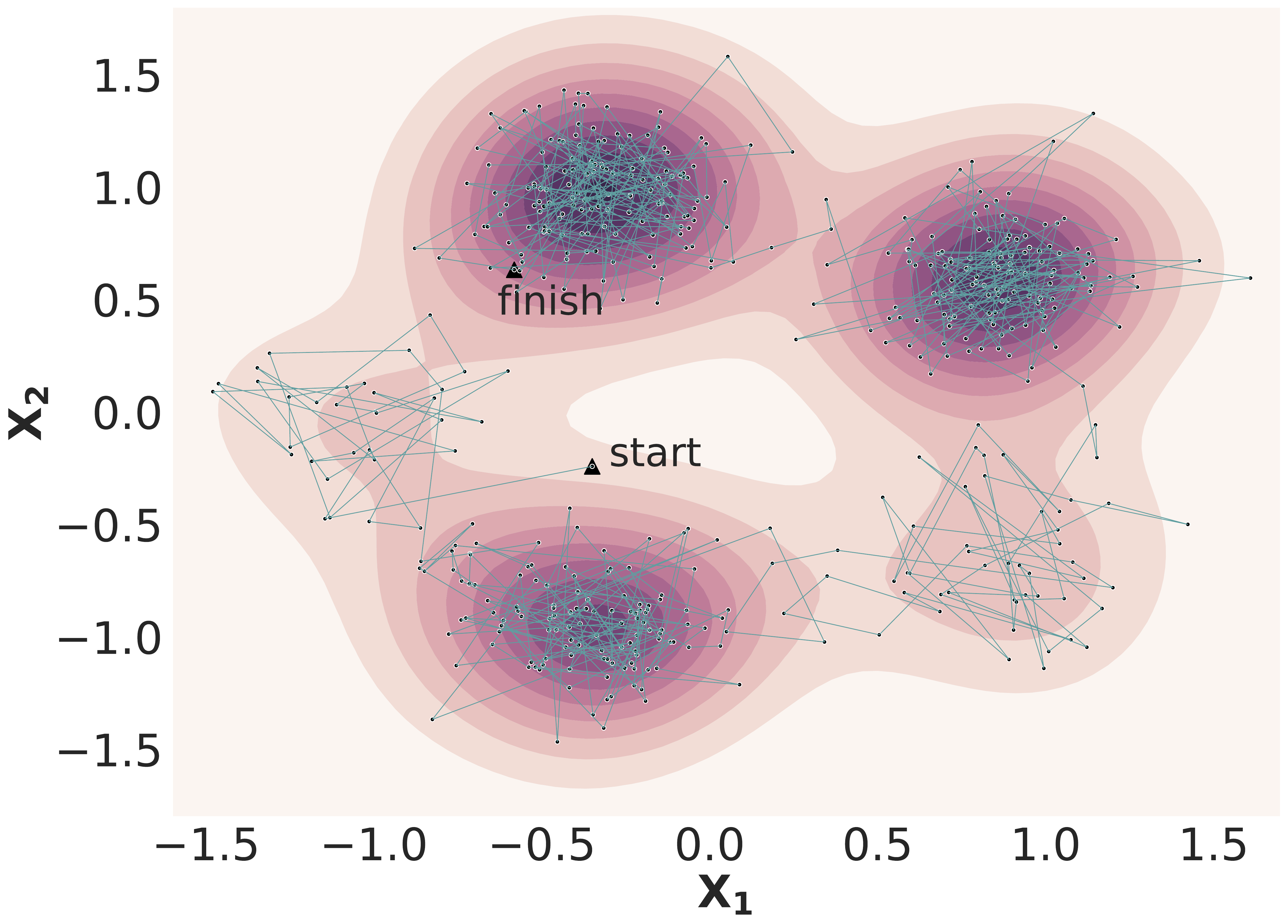}
    \caption{An example of a trajectory of a particle in the 5-well potential. Points indicate the positions of the particle at time $t$ and blue lines show the movement of the particle from time point $t$ to $t + 1$.}
    \label{fig:figure3}
\end{figure}

In order to obtain a time-evolving graph $\mathbb{G}$ with metastability, we use a stochastic differential equation to generate a trajectory based on which a set of time-snapshots of the graph $\mathbb{G}$ is then 
constructed.

Let us consider a particle in a 2-dimensional $s$-well potential given by the stochastic differential equation (SDE):
\begin{equation}
    dX_t = -\nabla F(X_t)dt + \sqrt{2\beta^{-1}}dW_t,
    \label{eqn:sde}
\end{equation}
with the potential 
\begin{equation*}
    F(x) = \cos(s\arctan(x_1, x_2)) + 10\Big(\sqrt{x^2_{1} + x^2_{2}} - 1\Big)^2.
\end{equation*}
See \citet{Klus2019KernelCC} for more details. Here, $s$ denotes the number of wells, since we assume that the number of wells defines the number of states in the time-evolving graph $\mathbb{G}$, the parameter $\beta$ is the inverse temperature and $W_t$ is a standard Wiener process. The particle stays in one of the wells for a relatively long time and then jumps to one of the neighboring wells. We consider one realization (trajectory) $\mathcal{S} \in \mathbb{R}^2$ of the stochastic process $X = \{X_t\}_{t = 0}^{L-1}$, where $L$ is the length of the trajectory.
An example of such a trajectory $\mathcal{S}$ is shown in Figure \ref{fig:figure3}, where the number of wells is $s = 5$ and $\beta = 0.05$.
Before generating a time-evolving graph $\mathbb{G}$, we cluster all points of $\mathcal{S}$ using \textit{k}-means in order to obtain the ground truth labels for time-snapshots of $\mathbb{G}$. Every synthetic benchmark data is based on this trajectory and constructed as follows.

\begin{figure}
    \centering
    \includegraphics[width=0.95\textwidth]{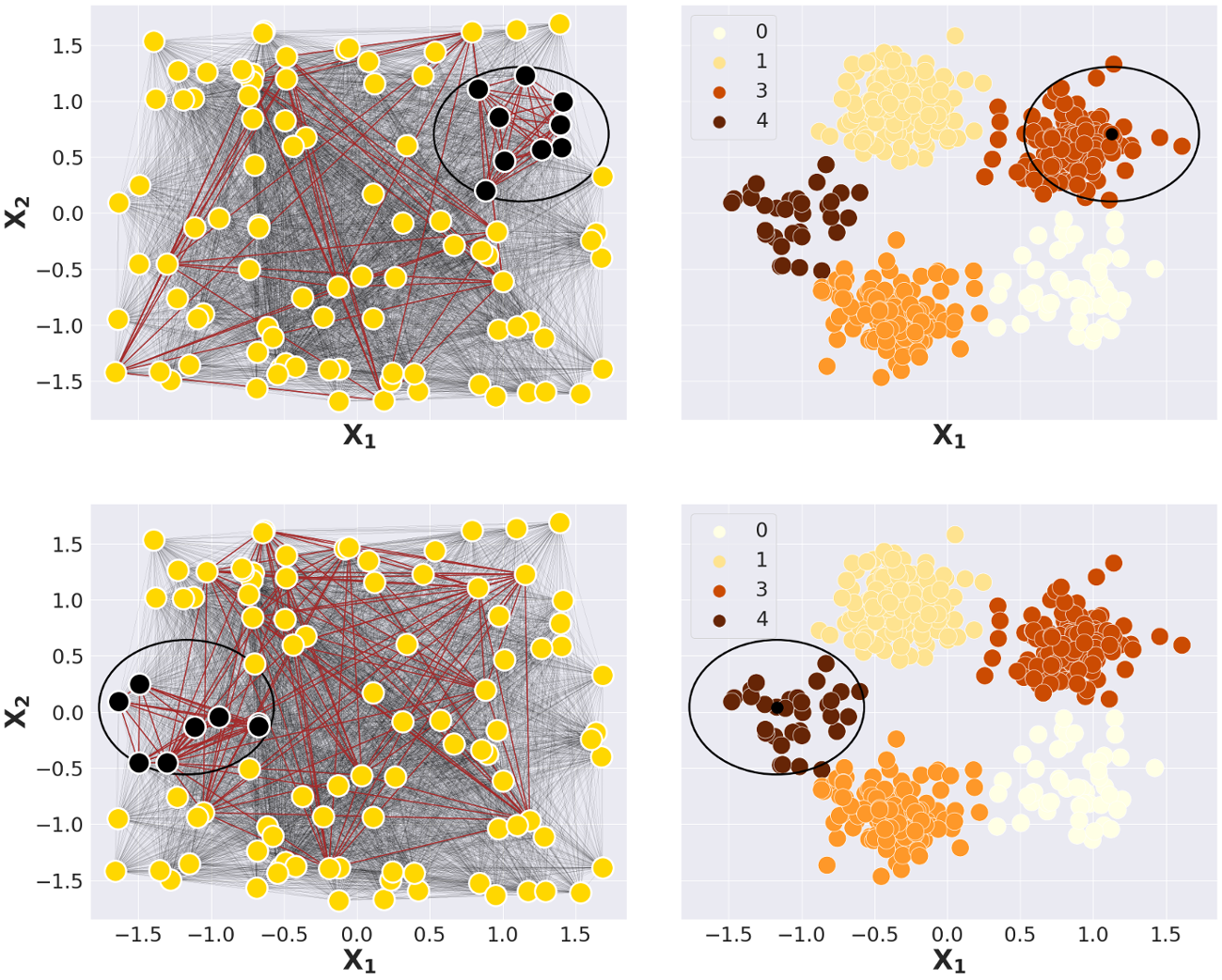}
    \caption{An illustration of our benchmark data at times \textbf{(a)} $t = 0$, \textbf{(b)} $t = 256$. In both \textbf{(a)} and \textbf{(b)}, the left images shows a time-snapshot $G_0$ and $G_{256}$ and the right images are points of the particle in the 5-well potential, which are clustered into 5 sets with \textit{k}-means. Edges in red color are removed from the graph. Vertices in the circles are considered as patterns characterizing corresponding states.} 
    \label{fig:figure4}
\end{figure}

The construction of the time-evolving graph $\mathbb{G} = \{G_0, ..., G_{T-1} \}$ can be described by a three-step process. In the first step, the trajectory $\mathcal{S} = \{(x_1^{(i)} x_2^{(i)})\}_{i = 0}^{L-1}$ using SDE \eqref{eqn:sde} is generated. We consider the case where the number of time points $T$ in $\mathbb{G}$ is equal to the length $L$ of $S$, we will then denote them both by $T$. In the second step, we choose the number of vertices $n$ and  assign positions $(a_j, b_j) $, $j=0, \dots, n - 1$ to each vertex $v \in V(G_t)$ in a Cartesian coordinate system. The number of vertices $n$ and their positions will be the same for each $G_t \in \mathbb{G}$, $t = 0, \dots, T-1$. We use the uniform distribution to generate random points $(a_j, b_j)$ such that $(a_j, b_j) \sim \mathcal{U}_{[-2, 2] \times [-2, 2]}$. Finally, in the third step of the construction process, we generate temporary patterns in the structure of the time-evolving graph such that it exhibits metastable behavior in the following way. At each time point $t \in \{0, \dots, T-1\}$, we draw a circle around the point $(x_1^{(t)}, x_2^{(t)}) \in \mathcal{S}$ with radius $r$. We choose the radius $r$ as the average of the radii of each cluster in $\mathcal{S}$ and $r$ is the same for each $t$. Each time-snapshot $G_t$ is first set to be a complete graph. We define temporal patterns, which characterize each state of $\mathbb{G}$, by removing all edges between vertices that are inside the current circle. In order to add noise to the data we also remove edges outside the circle with the  \textit{out-state} probability. An example of the benchmark data is shown in Figure \ref{fig:figure4}.


\section{Experiments and Results}
\label{sec:experiments}

We illustrate the efficacy of \textbf{graphKKE} proposed in Section \ref{sec: graphKKE} on the benchmark dataset and a real-world dataset with an artificial signal. We will show that our method is capable of learning the embedding of the time-evolving graph maintaining all dynamic properties in such way that it is possible to detect the metastable states in the low-dimensional space. Besides the experiments with benchmark and real-world datasets, we compare our method with several state-of-the-art approaches for graph clustering.


\subsection{\large{Experiments with synthetic data}}
\label{sec:exper_b_data}

\paragraph*{\textbf{Experimental setup.}} In order to test the performance of the method proposed in Section \ref{sec: method} and compare the result to other baselines models, we generate the synthetic data described in Section \ref{sec:data} with different configurations of interest such as the number of vertices $n$, the number of time steps $T$, and the number of states $s$. The datasets are summarized in Table \ref{tab:table1}. For each dataset we set the out-state probability to $0.1$. We apply \textbf{graphKKE} with the Weisfeiler--Lehman graph kernel with number of iterations $h = 1$ and regularization parameter $\eta=0.1$. In order to have ground truth labels/states of $\mathbb{G}$, we apply \textit{k}-means clustering to the SDE trajectory $\mathcal{S}$. For the Weisfeiler--Lehman kernel, the initial set of vertex labels $l_0$ is defined to be $\{0, 1, 2, \dots, n\}$.

\begin{figure}
    \centering
    \includegraphics[width=0.7\textwidth]{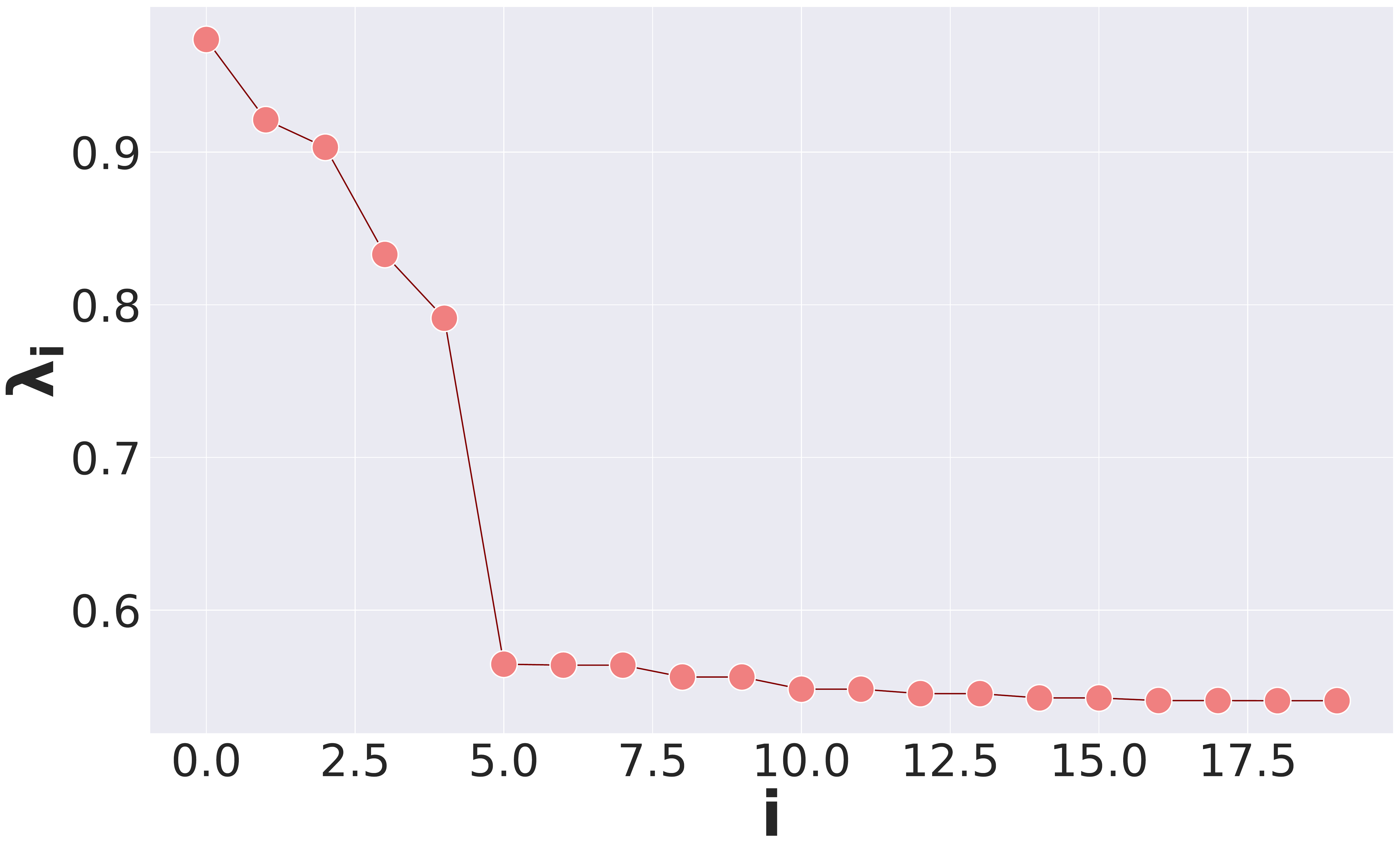}
    \caption{The eigenvalues of the Koopman operator approximated by \textbf{graphKKE} for 5DynG-100 dataset. The large spectral gap after the fifth eigenvalue reveals the presence of the 5 metastable states in the dataset.} 
    \label{fig:figure5}
\end{figure}

\begin{figure}
    \centering
    \includegraphics[width=0.8\textwidth]{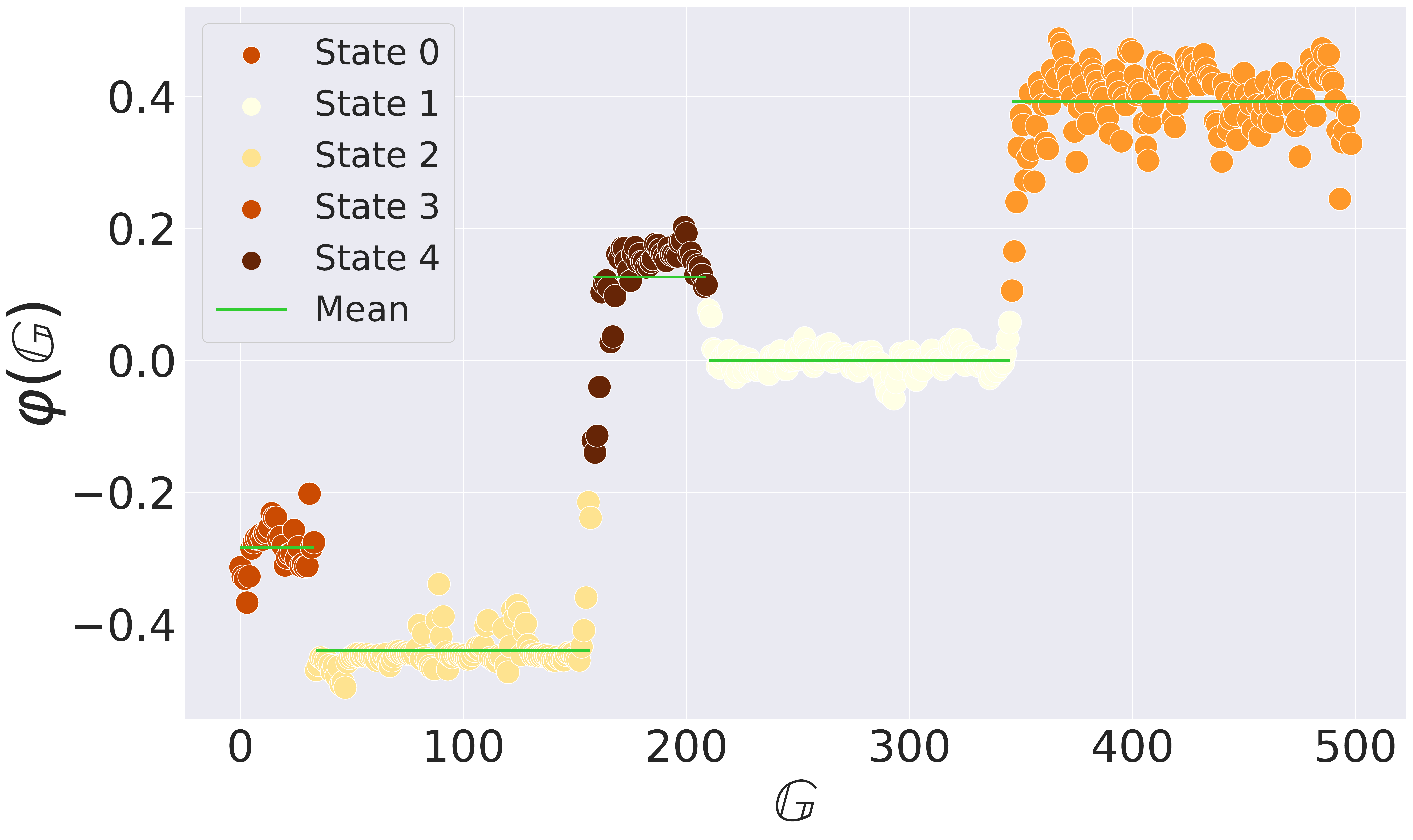}
    \caption{An illustration of the embedded time-series data, where $\varphi(\mathbb{G}) = \sum_{i=0}^4 c_i\varphi_i(\mathbb{G})$ and $\varphi_i(\mathbb{G})$ are dominant eigenfunctions of the Koopman operator approximated by graphKKE. There are $5$ distinct level sets corresponding to the $5$ metastable states.} 
    \label{fig:figure_new}
\end{figure}

\paragraph*{\textbf{Results and Analysis.}} We visualize the result only for the 5DynG-100 dataset. The eigenvalues of the Koopman operator approximated with \textbf{graphKKE} are shown in Figure \ref{fig:figure5}. A spectral gap after the fifth eigenvalue indicates that the time-evolving graph $\mathbb{G}$ contains $s=5$ metastable states and $\mathbb{G} = \mathbb{G}_0 \cup \dots \cup \mathbb{G}_{4}$. Since all information about the long-term behavior of the time-evolving graph is contained within the eigenfunctions of the Koopman operator associated with $s$ dominant eigenvalues close to 1 (in our case $s=5$), the embedding dimension $m$ is defined by the number of these eigenfunctions. Thus, for 5DynG-100 dataset each time-snapshot of $\mathbb{G}$ is embedded into a new vector space $\mathbb{R}^m$ with $s=m=5$. An illustration of the embedded time-series data $\varphi(G)$, where $\varphi(\mathbb{G}) = \sum_{i=0}^4 c_i\varphi_i(\mathbb{G})$, is shown in Figure~\ref{fig:figure_new}. Here, we chose the coefficients $c_i$ in such a way that the $5$ metastable states can be easily distinguished. Now we are able to analyze the data further using its low-dimensional representation and, for example, to detect the location of metastable states or to predict the state at the next time point.

\begin{figure}
    \centering
    \includegraphics[width=0.65\textwidth]{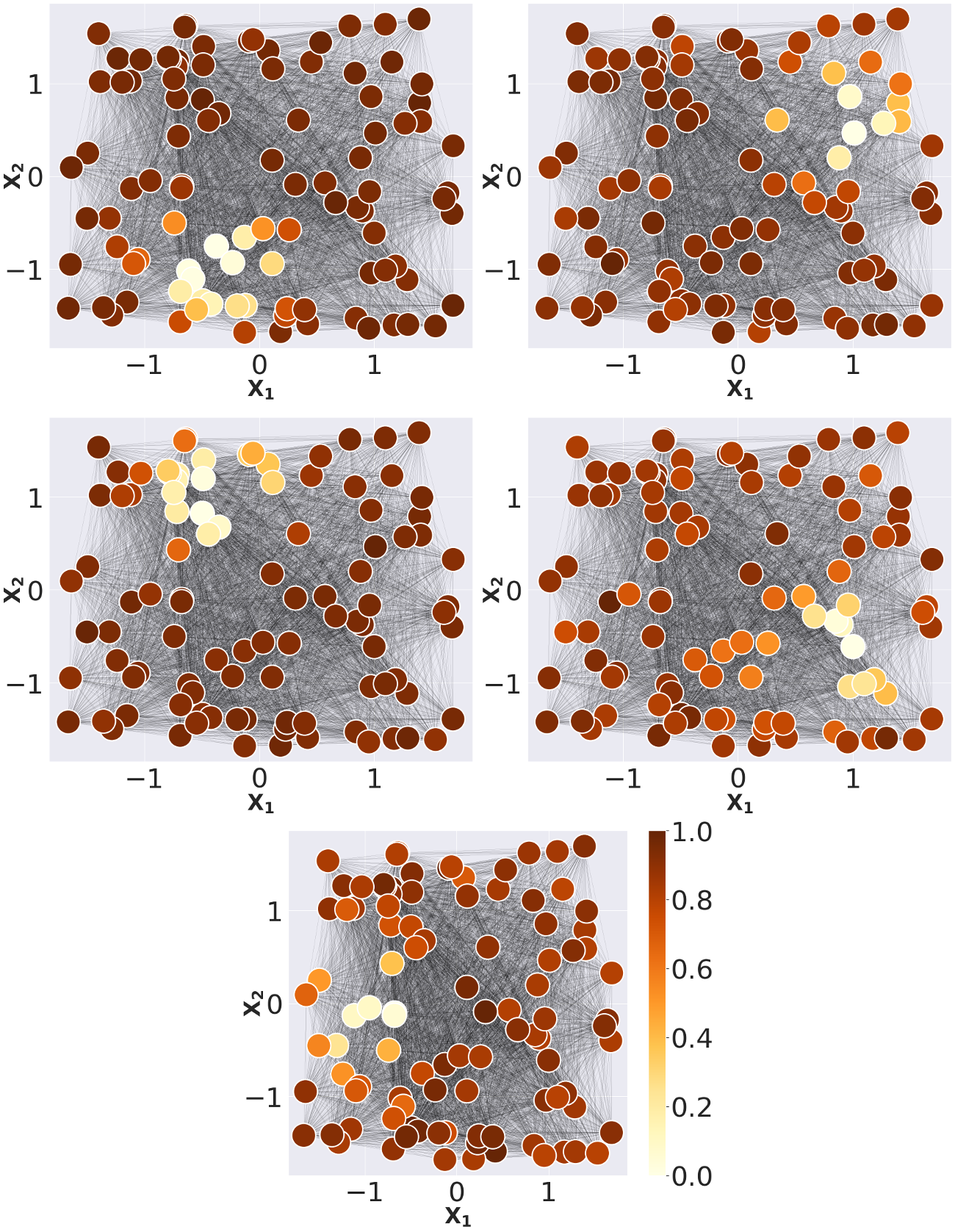}
    \caption{The result of detecting the metastable states using \textbf{graphKKE} for the 5Dyn-100 dataset. Each plot shows the average graph for each state, where color of the vertices reflects the average number of the edges. Vertices in brown have more edges than vertices in yellow and each state has a different cluster of vertices with fewer edges. Thus, the state-dependent graph exhibits a unique topology making it distinguishable from each other.}
    \label{fig:figure6}
\end{figure}

Applying \textit{k}-means to the eigenfunctions associated with the five dominant eigenvalues results in the five clusters. Since each state of the time-evolving graph is characterized by some common pattern in the topological structure, we average adjacency matrices of each state. Thus, if we have a time-evolving graph with s states $\mathbb{G} = \mathbb{G}_0 \cup \dots \cup \mathbb{G}_{s-1}$ and $\{\mathbb{A}_0, \dots, \mathbb{A}_{s-1}\}$ is a set of corresponding subsets of adjacency matrices, then
\begin{equation*}
    \mathbb{A}_{i}^{avg} = \frac{1}{|\mathbb{A}_i|} \sum_{j=0}^{|\mathbb{A}_i| - 1} A_i^j,
\end{equation*}
where $A_i^j \in \mathbb{A}_i, i = 0, \dots, s-1$. Each average adjacency matrix $\mathbb{A}_i^{avg}$ is associated with the average graph $\mathbb{G}_i^{avg}$.

Figure \ref{fig:figure6} illustrates the graphs of each state, where vertices are colored according to their degrees of the average graph $\mathbb{G}_i^{avg}, i = 0, \dots, s- 1$. 

Our approach is capable of capturing common temporal patterns in the topological structure of the time-evolving graph with metastability. Consequently, it can learn a meaningful embedding of the time-evolving graph and preserve states in a low-dimensional space.


\begin{table}
    \footnotesize
    \caption{\normalsize{Statistics of each dataset used in this paper.}}
    \centering
    \begin{tabular}{ccccc}
    \hline
    \textbf{Name} & \textbf{\#Vertices} & \textbf{\#Edges (avg.) $\pm$ std.} & \textbf{\#Time steps} & \textbf{\#States} \\[1ex]        
    \hline       
    5DynG-100 & 100 & 4851$\pm$43.68 & 500 & 5\\
    5DynG-200 & 200 & 19516$\pm$119.54 & 1000 & 5\\
    3DynG-300 & 300 & 44051$\pm$219.05  & 500 & 3 \\
    MovingPic & 919 & 10602$\pm$7266.39 & 658 & 2 \\
    CholeraInf & 96  & 106$\pm$41.28 & 34 & 2 \\
    \hline     
    \end{tabular}
    \label{tab:table1}
\end{table}

\subsection{\large{Experiments based on realistic data}}
\label{sec::movingpic_expriment}

In this experiment, we apply \textbf{graphKKE} to analyze a microbiome dataset, called \emph{MovingPic} coming from \citet{MovingPicture}, where one male and one female were sampled daily at three body sites (gut, skin, and mouth) for 15 months and for 6 months, respectively. As a feature matrix, the OTU table $D \in \mathbb{N}^{T \times p}$ is used, where $T$ is the number of time points and $p$ is the number of OTUs. The operational taxonomic units (OTUs) are defined as groups of closely related microbes or bacteria species.

We use the microbiome profile only from the skin and since the data does not have any perturbations such as antibiotics exposure or diseases, we add an artificial noisy signal to the data in the following way. A practical justification for adding noise to the signal is that the human microbiome might react not only to major perturbations such as diseases or antibiotics exposure but also to some short-term daily fluctuations such as changing of lifestyle or stress. Moreover, the noise will be added to test the robustness of \textbf{graphKKE}. Let $d_i= [d_i^0, d_i^2, \dots, d_i^{T-1}]$ be the $T$-dimensional column vector of OTU counts of the $ith$ species. OTUs with less than $30\%$ of total reads are removed from the matrix $D$. We randomly choose 100 OTUs that are used to add the noisy signal. The vector of length $T$ is constructed using a sine wave function:

\begin{equation*}
    z = R \cdot \sin\left(\frac{2\pi t}{\omega}\right)
\end{equation*}
and then for each $i, i = 0, \dots, 100$, we compute new OTU counts $d_i$,
\begin{equation*}
    d_i = d_i + \max(0, z + \epsilon \cdot w \cdot z),
\end{equation*}
where $w \sim \textit{Normal}(0, 1)$ and $\epsilon$ is the level of Gaussian noise. We set $\epsilon$ to one of $\{0, 0.05, 0.3\}$. 

The next step is the construction of a time-evolving graph. Let $d^t = [d^t_1, d^t_2, ..., d^t_p]$ be the $p$-dimensional row vector of OTU counts at time point $t, t = 0, ..., T-1$. The raw OTU counts are typically normalized by the total cumulative count $c^t = \sum_{i = 1}^{p} d_i^t$ in order to account for the different sequencing depth \citep{MetaNN}. Thus, the normalization of $d^t$ by the total cumulative count results in the relative abundance vector:
\begin{equation*}
    x_t = \Big[\frac{d^t_1}{c^t}, \frac{d^t_2}{c^t}, \dots, \frac{d^t_p}{c^t}\Big]
\end{equation*}
for each time point $t, t = 0, \dots, T-1$.
The time-snapshots of the time-evolving graph $\mathbb{G} = (G_0, ..., G_{T-1})$ are then constructed as follows. First of all, we compute the Pearson correlation coefficient of each pair of OTUs $(d^i, d^j)$, with $i, j = 1, ..., p$ in order to define an initial co-occurrence graph. We choose a threshold of $0.5$ such that edges with the Pearson coefficient greater than $0.5$ or less than $-0.5$ are considered to be strongly correlated and remain in $G_0$. Edges with the Pearson correlation coefficient in the range $[-0.5; 0.5]$ are removed from the initial graph. Furthermore, to construct time-snapshots for each $t = 0, \dots, T-1$, we use the OTU counts. If the OTU count for the current vertex is zero, we remove edges connecting this vertex and its neighboring vertices. The statistics of the pre-processed data can be seen in Table \ref{tab:table1}.

Moreover, we define $\widehat{\mathbb{G}}_t = \mathbb{G}_{t+\tau}$. That is, for the chosen lag time $\tau = 1$, $\mathbb{G} = (G_0, \dots, G_{T-2})$ and $\widehat{\mathbb{G}} = (G_1, \dots, G_{T-1})$. From the two time-evolving graphs $\mathbb{G}$ and $\widehat{\mathbb{G}}$, we compute the Gram matrices $K_{\mathbb{G}\mathbb{G}}$ and  $K_{\mathbb{G}\widehat{\mathbb{G}}}$ using the Weisfeiler--Lehman kernel, where the number of iterations is set to $h=1$, and the regularization parameter to $\eta=0.9$.

\begin{figure}
    \centering
    \includegraphics[width=0.8\textwidth]{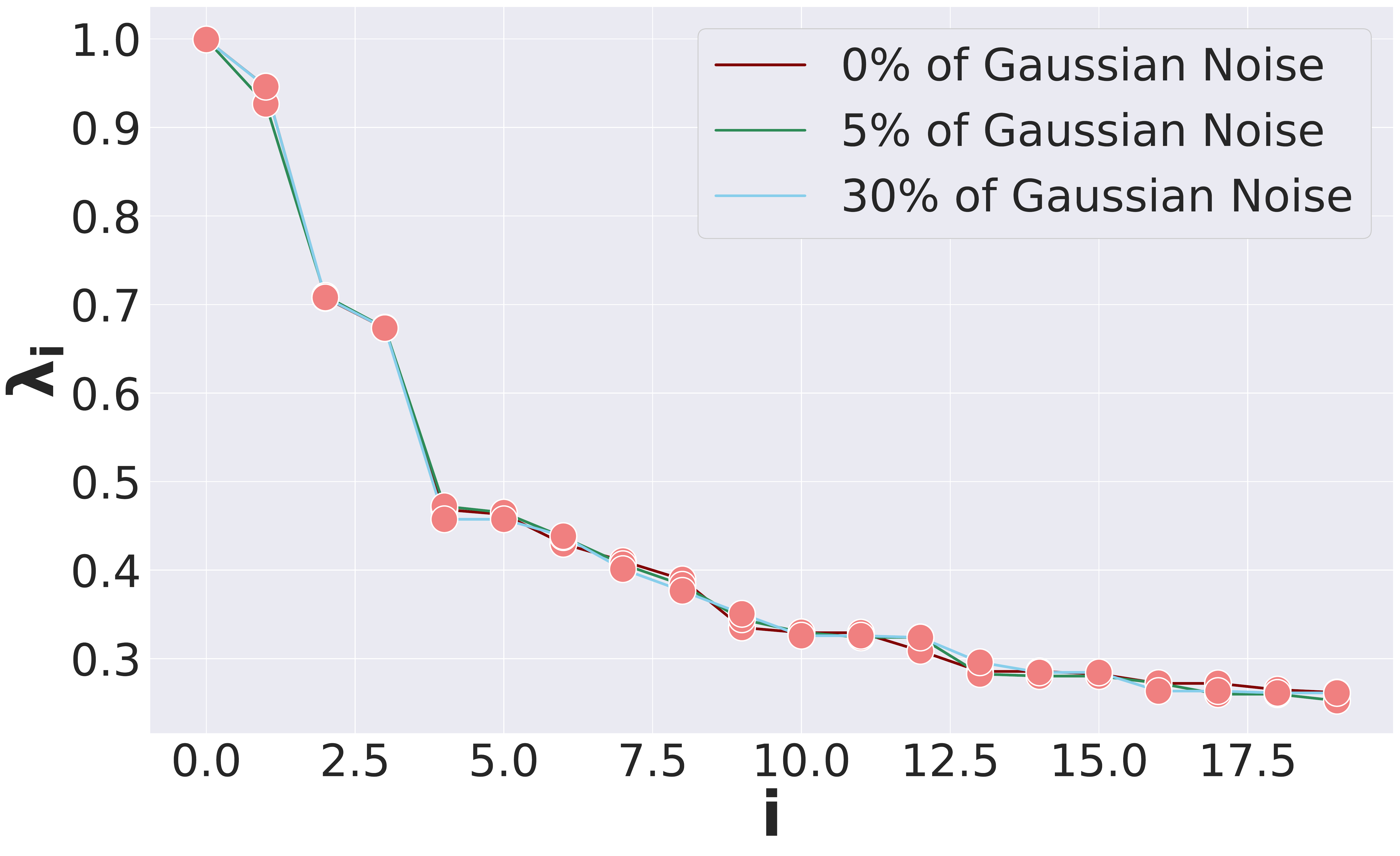}
    \caption{The eigenvalues of the Koopman operator approximated by \textbf{graphKKE} for different percentages of Gaussian noise added to the MovingPic dataset.}
    \label{fig:figure8}
\end{figure}

\paragraph*{\textbf{Results \& Analysis.}}
The eigenvalues detected by \textbf{graphKKE} for different percentages of Gaussian noise are shown in Figure \ref{fig:figure8}. The gap after the second eigenvalue and the values of these eigenvalues close to 1 imply the presence of two states in the time-evolving graph $\mathbb{G}$. The spectral gap after the forth eigenvalue indicates the presence of four states but we are not aware of the biological interpretations of the second two states since the original study does not mention any potential perturbations. The experiment also shows that \textbf{graphKKE} is robust to the noise in the data. In order to find the location of the states, we cluster time-snapshots into two states using \textit{k}-means applied to the two normalized eigenfunctions associated with two dominant eigenvalues with the number of clusters set to 2.

The following experiment will demonstrate whether the detected states in the benchmark and the real-world datasets correspond to the ground truth labels. Moreover, we will show that \textbf{graphKKE} outperforms other methods for learning the embeddings of time-evolving graphs.


\begin{table}
    \footnotesize
    \caption{Hyperparameters for \textbf{graphKKE} used in the comparative analysis in Section \ref{sec:comp_analysis}, where $\sigma$ is the bandwidth, $h$ the number of iterations, and $\eta$ the regularization parameter.}
    \centering
    \begin{tabular}{cccc}
    \hline
    \textbf{Dataset} & \textbf{$\sigma$} & \textbf{$h$} & \textbf{$\eta$} \\[1ex]
    \hline 
    5DynG-100 & 10 & 1  & 0.1  \\ 
    5DynG-200 & 100 & 1 & 0.5 \\ 
    3DynG-300 & 100 & 1 & 0.1  \\ 
    MovingPic & 100 & 1 & 0.5 \\ \hline
    \end{tabular}
    \label{tab:table2}
\end{table}

\subsection{\large{Comparative analysis}}
\label{sec:comp_analysis}

\paragraph*{\textbf{Experimental setup.}} The goal of this experiment is to compare \textbf{graphKKE} to several state-of-the-art representation learning and graph clustering approaches using benchmark and real-world datasets. The proposed approach with two different graph kernels --- Gaussian and Weisfeiler--Lehman kernels --- is compared with graph2vec \citep{graph2vec} and the original WL kernel \citep{WL}. The main idea of graph2vec is explained in Section~\ref{sec:intro} and the WL kernel is discussed in Section~\ref{sec: graph_kernel}. Since the analysis is done for the graph clustering task, we apply \textit{k}-means to the resulting embedding vectors of every approach. The embedding dimensions of $\{5, 64, 128, 1024\}$ were chosen for graph2vec. The hyperparameters of \textbf{graphKKE} were chosen empirically\footnote{The combinations of hyperparameters with the biggest spectral gap were used.} and can be seen in Table~\ref{tab:table2}. The choice of $\sigma$ for the Gaussian kernel is critical for the performance of \textbf{graphKKE}. The optimal choice of $\sigma$ is beyond the scope of this paper (for details see \citet{ChoiceSigma}). For the MovingPic dataset, the level of Gaussian noise is set to $0.05$ in this experiment.

\paragraph*{\textbf{Evaluation Metric.}} In order to assess the results of the clustering of the embedding vectors for all approaches, the Adjusted Rand Index (ARI) is used. Higher ARI corresponds to greater accuracy in correctly identifying the ground truth labels/states.

\paragraph*{\textbf{Results \& Analysis.}} The graph clustering results for all datasets using \textbf{graphKKE} and other state-of-the-art methods are presented in Table \ref{tab:table3} (experimental datasets). For graph2vec the embedding dimension of 5 was used as a dimension with the best ARI to compare its result with the results of other approaches. We observe that both graph2vec and WL kernel perform poorly on the benchmark and real-world datasets. One reason of the poor embedding is that these two methods do not take into account the time information which is crucial in time-evolving graphs with metastability. 

Additionally, this experiment shows that the detected metastable states using the embedding of \textbf{graphKKE} correspond exactly to the ground truth labels. In the benchmark data, the ground truth labels are the labels of the \textit{k}-means clustering of the trajectory $S$. In the case of the MovingPic dataset, the ground truth labels correspond to the time period when the sine wave function of the artificial signal is zero (label 0) or greater than zero (label 1). 

\begin{table}
    \footnotesize
    \caption{Adjusted Rand Index (ARI) for the comparative analysis on the graph clustering task in Section \ref{sec:comp_analysis}. Higher ARI corresponds to greater accuracy in correctly identifying the ground truth states. It can been seen that the combination of \textbf{graphKKE} with Weisfeiler-Lehman kernel outperforms other methods.}
    \begin{tabular}{ccccc}
    \hline
    \textbf{Dataset} & \textbf{graph2vec} & \textbf{WL kernel} & \textbf{graphKKE+WL kernel} & \textbf{graphKKE+Gaussian kernel} \\[1ex]
    \hline
    & \multicolumn{4}{c}{\textit{Experimental datasets}} \\ \hline
    5DynG-100 & 0.49 & 0.36 & \textbf{0.99} & 0.96 \\
    5DynG-200 & 0.20 & 0.49 & \textbf{0.92} & 0.87 \\
    3DynG-300 & 0.22 & 0.40 & \textbf{0.96} & 0.94 \\
    MovingPic & 0.42 & 0.56 & \textbf{1} & 0.99\\ \hline
    
    & \multicolumn{4}{c}{\textit{Real-world dataset}} \\ \hline
    CholeraInf & 0.29 & 0.66 & \textbf{0.88} & 0.87 \\ \hline
    \end{tabular}
    \label{tab:table3}
\end{table}


\section{Application to microbiome data}
\label{sec:microbiome_d}
Having studied the performance of \textbf{graphKKE} on benchmark datasets and the real-world dataset with the artificial signal, we now describe the application of our \textbf{graphKKE} approach to the microbiome data. Such data is more challenging than the benchmark data because the real-world data generating process is more complex and also contains noise.

\paragraph*{\textbf{Background.}} The microbiome data, which we will analyze in this section comes from a study about recovery from Vibrio cholerae infection \citep{CholeraInfOriginal}. Fecal microbiota was collected during acute diarrhea and recovery periods of cholera in a cohort of seven Bangladeshi adults. In our experiments, we chose one patient, since there is variation in the constituents of the gut microbiota among individuals \citep{DiffCompositionMicrobiome} and thus, it can bias the result of detecting the metastable states such as diarrhea and recovery periods. The pre-processed OTU table were obtained from \citet{CholeraInf}. The aim is to determine if there are metastable states in this data and if possible, the number of metastable states and their locations.

\begin{figure}
    \centering
    \includegraphics[width=0.7\textwidth]{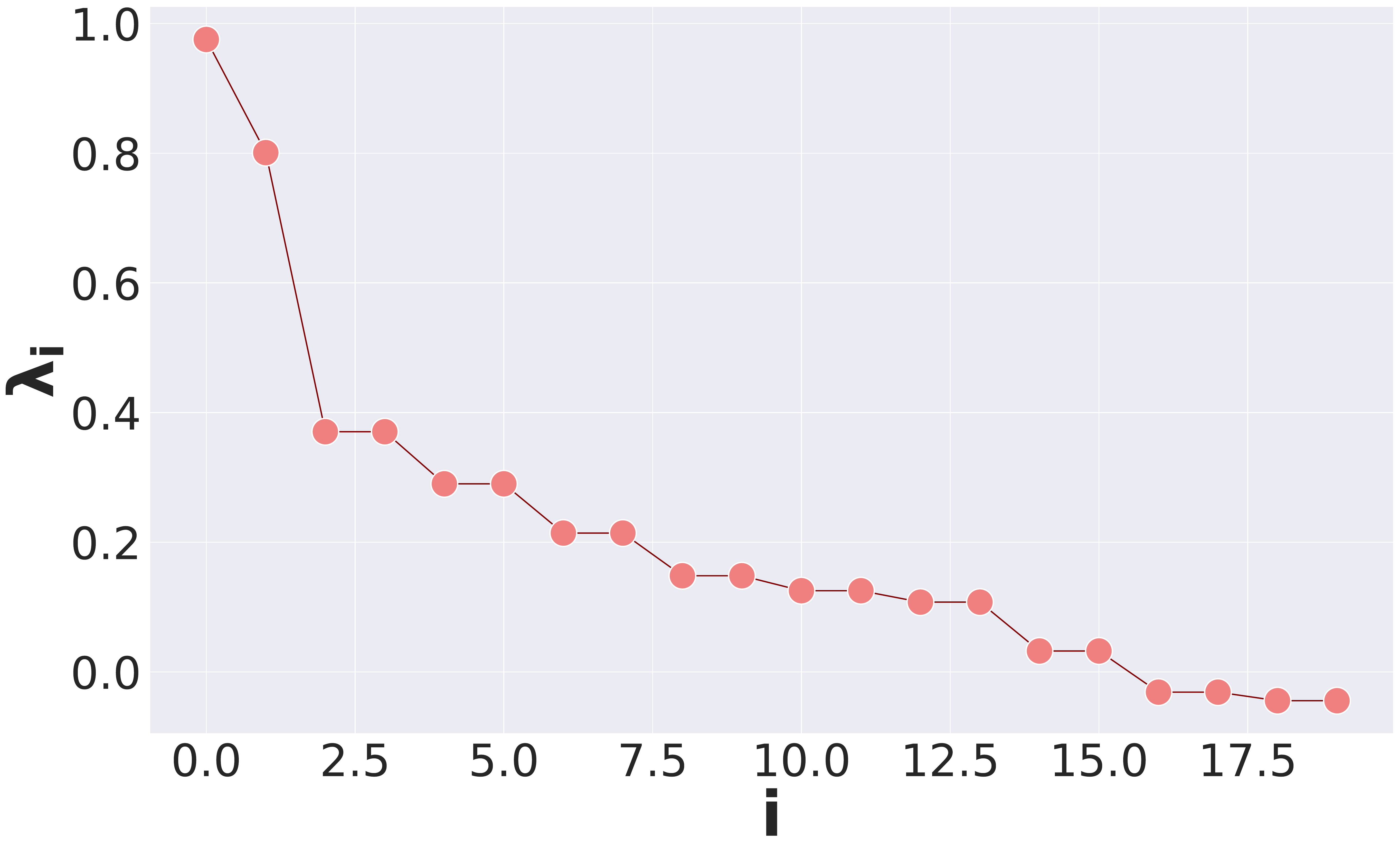}
    \caption{The large spectral gap after the second eigenvalue of the Koopman operator approximated by \textbf{graphKKE} indicates that the cholera infection dataset can be divided into two metastable states.} 
    \label{fig:figure9}
\end{figure}

The time-evolving graph from the given OTU table is constructed in the same way as for the MovingPic dataset using the relative abundance vector and Pearson correlation coefficients. In the real-world microbiome dataset, perturbations do not always shift OTU counts to zero. Therefore, the question how to properly construct time-evolving graphs such that both metastable behavior and associations between microbes are taken into consideration need to be considered in future work. 

We apply \textbf{graphKKE} using the Weisfeiler--Lehman graph kernel. We set the number of iteration to $5$ and the regularization parameter to $0.1$. 

\begin{figure}
    \centering
    \includegraphics[width=0.8\textwidth, height=0.95\textwidth]{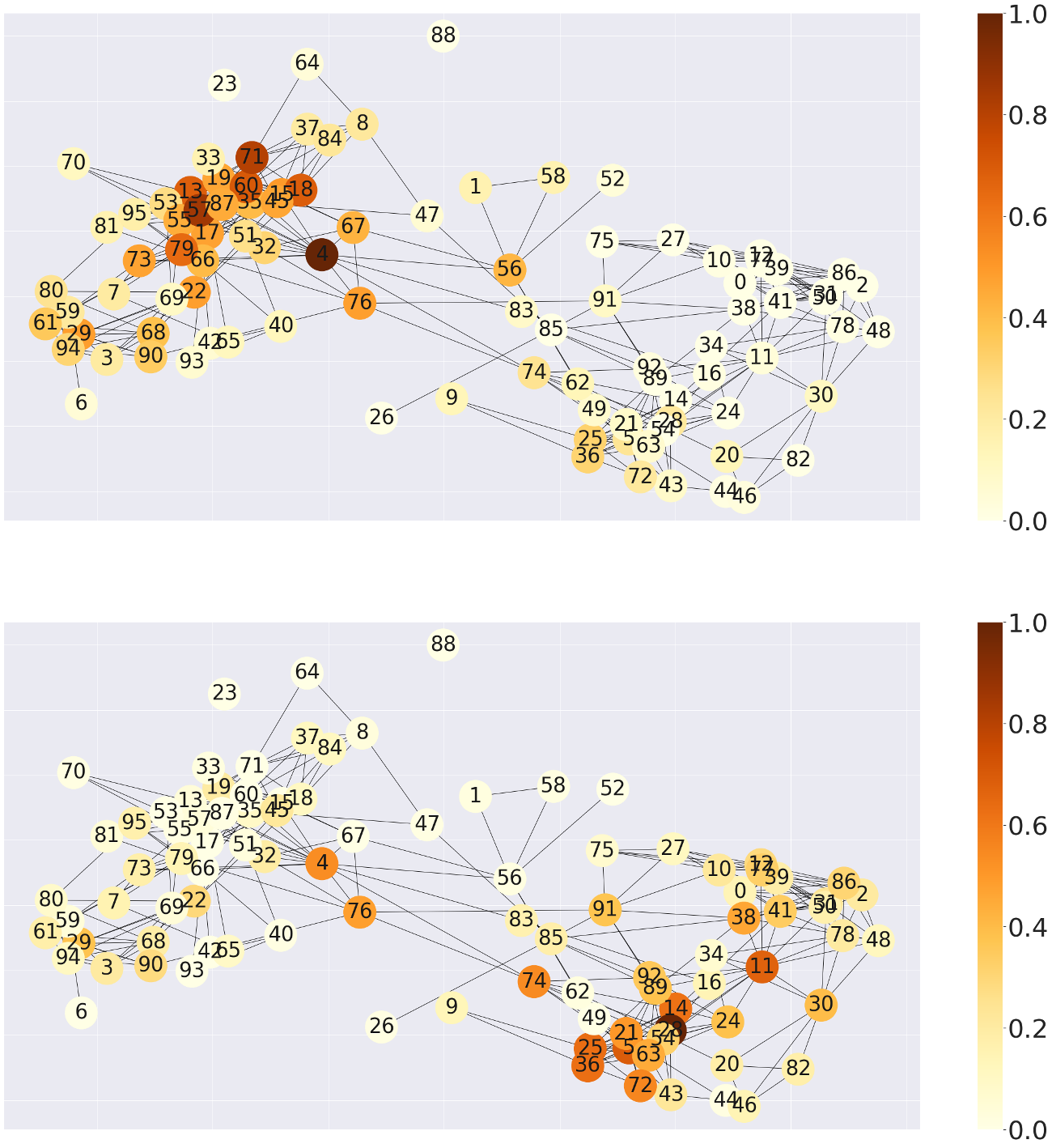}
    \caption{The same interaction graph constructed with Pearson correlation coefficients for two different states detected using \textbf{graphKKE}. Vertices are colored according to the average number of edges: \textbf{(a)} period of cholera infection. \textbf{(b)} period of recovery. Brown color implies that particular species (vertices) have more association interactions than species (vertices) in yellow.} 
    \label{fig:figure10}
\end{figure}

\paragraph*{\textbf{Results \& Analysis}}
The resulting eigenvalues are shown in Figure \ref{fig:figure9}. Two dominant eigenvalues close to 1 implies that the time-evolving graph $\mathbb{G}$ contains two metastable states and further in the paper we will show that these two metastable states correspond to the ground truth infection/recovery periods of the dataset. Moreover, the eigenfunctions associated with these two dominant eigenvalues contain all information about the long-term behavior of the time-evolving graph $\mathbb{G}$ and using them as a low-dimensional representation we can further analyze the cholera dataset with the aid of time-series methods which work with vector-structured data. For example, one can cluster the data into two clusters, predict the state at the next time point or we can find the probability of $\mathbb{G}$ returning to the diarrhea state if a person continues living in this area.

We will focus on detecting metastable states utilizing the low-dimensional representation (dominant eigenfunctions). Applying a clustering method such as \textit{k}-means to the two dominant eigenfunctions, we can find the location of metastable states in $\mathbb{G}$. Moreover, in order to estimate whether the resulting embedding maintains the dynamics of the time-evolving graph, we will compare the metastable states, which we obtained by clustering the two dominant eigenfunctions, with the initial time periods of diarrhea and recovery. The ARI is shown in Table~\ref{tab:table3} (real-world dataset).

After clustering eigenfunctions into two states, we can compare the topological structures of time-snapshots of these states. We compute the average adjacency matrices in each state as discussed in Section~\ref{sec:exper_b_data}. The result is shown in Figure~\ref{fig:figure10}. We see that, depending on the state, different clusters of vertices have different degrees. This is due to the fact that the cholera infection causes marked shifts in the microbiome composition. The biological meaning of these clusters and how they are related to the healthy/ill state are open questions and need to be analyzed in future work.

This result shows that the embedding of the time-evolving graph $\mathbb{G}$ simplifies the analysis of graph-structured time-series data and can be used to extract crucial properties of the graph that make the time-series graph undergo transitions from one state to another.

\section{Discussion \& Conclusion}

The large variety of species and complex interactions in the microbiome makes it challenging for researchers to analyze the responses of the microbiome to different perturbations such as diseases or antibiotic exposures and its influence on the human health. However, most studies aiming at understanding these dynamics are primarily focused on statistical constitution analysis omitting more complex interactions that can be described as a time-evolving graph. One solution is to represent each time-snapshot of the time-evolving graph as a fixed-length feature vector. Many existing approaches learn the embedding either of the static graphs or of the substructures such as nodes, edges, or subgraphs, whereas for some system it is of great importance to embed the entire time-snapshots of the time-evolving graph into a low-dimensional space preserving the global temporal mechanisms such as metastability.

In this paper, we introduced an unsupervised approach (i.e., class labels of single time-snapshots are not required to learn the embedding) for learning a mapping that embeds time-snapshots of a time-evolving graph exhibiting metastable behavior as points in a low-dimensional vector space.  Our experiments on synthetic benchmark and real-world data show that our approach is capable of learning a low-dimensional representation of the time-evolving graph that preserves the metastable behavior. This embedding can then be clustered in order to split individual time-snapshots of the time-evolving graph into states. Moreover, one can also analyze the dynamics occurring in the time-evolving graph (e.g., the probability of jumping from one state to another or the probability that the graph will return to one of the states) and apply different machine learning techniques. Since we are dealing with graph-structured data, which usually represents the interactions between objects, we can extract structural information pertaining to particular states. The latter is beneficial in the case of biological interactions such as microbiome data, where it is crucial to understand the differences between states (e.g., healthy/ill). To this end, experimental results have shown that our approach can outperform several state-of-the-art methods for representation learning of graphs. For instance, the comparative analysis has shown that applying only Weisfeiler--Lehman kernel to the time-evolving graph is not sufficient to capture the underlying dynamical graph patterns and consequently, to detect the metastable sets.

We have shown that graph kernels are not only a powerful tool for analyzing static graphs but also for analyzing time-evolving graphs. The transfer operator approach in combination with graph kernels yields a method capable not only of extracting structural information in each time-snapshot of the time-evolving graph but also of identifying the evolution patterns, which may exist in time-evolving graphs with metastability over long periods of time.

\section*{Acknowledgements}

This work was supported by the German Ministry for Education and Research (BMBF) within the Berlin Big Data Center and the Berlin Center for Machine Learning (01IS14013A and 01IS18037J) and the Forschungscampus MODAL (project grant 3FO18501) and funded by the Deutsche Forschungsgemeinschaft (DFG, German Research Foundation) under Germany's Excellence Strategy – The Berlin Mathematics Research Center MATH+ (EXC-2046/1, project ID: 390685689).

\bibliographystyle{plainnat}
\bibliography{article}

\end{document}